\documentstyle[12pt]{article}
\newcommand{\be}{\begin{equation}}
\newcommand{\ee}{\end{equation}}
\newcommand{\bea}{\begin{eqnarray}}
\newcommand{\eea}{\end{eqnarray}}
\newcommand{\nn}{\nonumber \\}
\newcommand{\p}[1]{(\ref{#1})}
\newcommand\tb{\bar\theta}

\newcommand{\Db}{\bar{D}}
\newcommand{\cD}{{\cal D}}
\begin{document}
\vskip1cm
\centerline{\large\bf 1/4 PBGS and
  Superparticle Actions}
\vskip1cm
\centerline{\bf F. Delduc${}^{\, a}$, E. Ivanov${}^{\, b}$, S.
Krivonos${}^{\, b}$}
\bigskip
\centerline{$^a$ {\it Laboratoire de Physique, }}
\centerline{\it  Groupe de Physique 
Th\'eorique ENS Lyon}
\centerline{\it 46, all\'ee d'Italie, F - 69364 - Lyon CEDEX 07}
\bigskip
\centerline{$^b${\it Bogoliubov Laboratory of Theoretical Physics, 
JINR,}}
\centerline{\it 141 980, Dubna, Moscow Region, Russian Federation}

\vskip1cm

\begin{center}
{\begin{minipage}{4.2truein}
                 \footnotesize
                 \parindent=0pt
We construct the worldline superfield
massive superparticle actions which preserve $1/4$ portion of
the underlying higher-dimensional supersymmetry. 
We consider  the cases of $N=4 \rightarrow N=1$ and  $N=8 \rightarrow N=2$
partial breaking. In the first case we  present the corresponding
Green-Schwarz type target superspace action with one $\kappa$-supersymmetry.
In the second case we find out two possibilities, one of which is a direct
generalization of the $N=4 \rightarrow N=1$  case, while another is
essentially different.
\par\end{minipage}}\end{center}
 \vskip 2em \par
\noindent{\bf 1. Introduction.}\hskip 1em
The most attractive feature of the
description of superbranes based on idea of the partial 
spontaneous breaking of global supersymmetry (PBGS)
\cite{hp} - \cite{ik} is the manifest off-shell realization of the
worldvolume supersymmetry. In this approach, the
physical worldvolume multiplets 
are interpreted as Goldstone superfields realizing spontaneous
breaking of the full brane supersymmetry group down to its unbroken
worldvolume subgroup. 
The invariant Goldstone superfields actions, after
passing to the component fields, coincide with 
the static gauge forms of the relevant Green-Schwarz (GS) type actions.

Until present, only the examples
of $1/2$  breaking of supersymmetry corresponding to the
standard  BPS  $p$-branes and D-branes were treated in the
literature on the PBGS. 
It is interesting to
extend the PBGS framework to the $1/4$ breaking  and other fractional
patterns (see, e.g.,  \cite{bdl}-\cite{gh} for the discussion 
of such options at the algebraic level).

In the present talk we
describe several  examples of the $1/4$  PBGS superfield actions in the
simplest case of massive superparticles, namely, the actions 
corresponding to the PBGS patterns $N=4\; \rightarrow \;N=1$ 
and $N=8 \;\rightarrow \;N=2$. In the first case we construct 
a worldline superfield action and show its equivalence to 
the GS-type target superspace action with one 
fermionic $\kappa$ symmetry. In the second case we find out two different 
models. For one of them we find the bosonic part of the worldline superfield 
action and the corresponding GS-type action with two $\kappa$ 
symmetries. A common feature of all the cases considered is that the 
algebras of their underlying spontaneously broken $d=1$ supersymmetries are 
dimensionally-reduced forms of $N=1$ and $N=2$ $D=4$ Poincar\'e  
superalgebras extended by tensorial central charges 
\cite{agit,ferrp,blstenz,gh}.
\vspace{0.3cm}

\noindent{\bf 2. N=4 $\rightarrow$ N=1 PBGS.} \hskip 1em 
Our goal here will be to construct a
$N=1, d=1$ superfield
action which respects three extra spontaneously
broken supersymmetries.
Thus, the minimal multiplet should include at least three $N=1$ fermionic
Goldstone superfields $\psi^i(t,\theta) (i=1,2,3)$
 given on $N=1, d=1$ superspace with the coordinates
 $\left\{ t,\theta \right\}, ( {\bar t}=-t,\bar\theta=\theta )$.
They should inhomogeneously transform under the broken supersymmetries. 
One can check that this requirement is met in a minimal way and the algebra
of transformations gets closed at cost of adding one additional
fermionic $N=1$ superfield $\Upsilon(t,\theta)$. The broken 
supersymmetry transformations read
\be \label{n4tr1}
\delta \psi^i = \epsilon^i\left(1
-\cD\Upsilon\right)-  \varepsilon^{ijk}\epsilon^j \cD \psi^k \;, \;
\delta\Upsilon=\epsilon^i \cD \psi^i~ \;,
\ee
where 
\be
{\cal D}=\frac{\partial}{\partial \theta}+\theta\partial_{t}\; ,\;
\left\{ \cD, \cD \right\} =2\partial_{t}~.
\ee
They form the following algebra\footnote{Hereafter, 
we deal with the algebras of the superfield variations. 
The superalgebras of the supercharges constructed by the PBGS actions 
following the N\"other procedure
are different: they inevitably include some constant central
charges which are crucial for evading \cite{{hp},{hlp}} the famous Witten's
no-go  theorem \cite{Witten}.} 
\be
\left\{ Q, Q \right\} = 2P\; , \;
\left\{ S^i, S^j \right\} = 2\delta^{ij}P\; , \;
\left\{ Q,S^i \right\} = 0\;. \label{susy4a}
\ee

We wish to have a superparticle model with the worldline scalar 
$N=1$ multiplets containing physical bosonic fields. Thus 
we are led to introduce bosonic superfields $v^i$
\be
\psi^i = \frac{1}{2}\cD v^i \;, \;
\left( {\bar\psi}{}^i=\psi^i\;, \; {\bar v}{}^i=-v^i \right) \;,
\ee
\be\label{n1tr2}
\delta v^i = -2\epsilon^i\left(\theta -\Upsilon\right)+
 \varepsilon^{ijk}\epsilon^j \cD v^k \;, \;
\delta\Upsilon=\frac{1}{2}\epsilon^i \partial_t v^i \; .
\ee
Due to  the explicit  presence of
$\theta$ in the transformations \p{n1tr2}, the anticommutators of
$Q$ and the spontaneously broken supersymmetry generators $S^i$
acquire active central charges $Z^i$ in the right-hand side: 
\be
\left\{ Q,S^i \right\} =2 Z^i~. \label{susy4}
\ee
The central charge generators act as pure shifts of $v^i$, suggesting
the interpretation of $v^i$ as Goldstone superfields
parametrizing transverse directions in a four-dimensional space
where $Z^i, P$ act as the translation operators.

Surprisingly, the superalgebra \p{susy4} cannot be interpreted as a
dimensional reduction of the standard $N=1$ Poincar\'e superalgebra
in $d=4$, with $Z^i, P$ being the components of full $4$- momentum.
One should proceed
not from the standard $N=1$, $D=4$ super Poincar\'e algebra, but from its
extension by tensorial central charges \cite{agit,ferrp,blstenz,gh}.
The generators $Z^i$ turn out to partly come from these central charges
and partly from the extra components of  $4$-momentum. Namely, 
\be
P={1\over 2}\,P_0~, \quad Z^1 = P_2~, \quad Z^2 = -P_1~, \quad Z^3 = 
{i\over 4}(\overline{T}_{22} - 
T_{22})~, 
\ee
where the original $N=1$, $D=4$ superalgebra is defined by the relations
\bea
&& \{Q_\alpha, \; \bar Q_{\dot \beta} \} =
2\,(\sigma^m)_{\alpha\dot \beta}P_{m}~, \nn
&& \{Q_\alpha, \; Q_{\beta} \} = 2\,T_{(\alpha\beta)}\;, \qquad
\{\bar Q_{\dot \alpha}, \; \bar Q_{\dot \beta} \} =
2\,\overline{T}_{(\dot \alpha\dot \beta)}~. \label{A2}
\eea

Let us construct invariant action for the
system under consideration.
The field $\Upsilon(v)$
is a good candidate for the Lagrangian density
\be\label{n4action1}
S_v= \int dt d\theta \Upsilon(v) \;,
\ee
in view of its transformation property \p{n4tr1}.
Then the question is how to covariantly express $\Upsilon $ in
terms of $\psi^i$ and, further, $v^i$. This can 
be done rather easily.

The most general ansatz for $\Upsilon$ is as follows
\bea
\Upsilon & = & \psi^i\cD\psi^i A + \psi^{2i}\psi^i_tB +\psi^{2i}\cD\psi^i
    \psi^j_t\cD\psi^j C \nn
&& +\psi^i\psi^i_t\psi^j\cD\psi^jE 
+ \psi^3\cD\psi^i_t\cD\psi^i F +\psi^3\psi^{2i}_t\cD\psi^i G~, 
\label{n1action2}
\eea
where $A,B,\ldots,G$ are as yet undetermined functions of $X$,
and we use the following notations
\be
\psi_t^i = \partial_t \psi^i\;, \;
\psi^{2i} =\varepsilon^{ijk}\psi^j\psi^k \;, \;
\psi^3= \varepsilon^{ijk}\psi^i\psi^j\psi^k~.
\ee
Now, using \p{n4tr1}, \p{n1action2} we can write $\delta\psi^i$ in
terms of $\psi^i$. Then we explicitly evaluate
$\delta\Upsilon$ and require it to be equal to $\epsilon^i \cD
\psi^i$ in accordance with the transformation law \p{n4tr1}.
After rather lengthy calculations we  get the system of {\it algebraic}
equations for the unknowns $A,\ldots, G$ \bea\label{coeff}
&&A=\frac{2}{1+\sqrt{1-4\cD\psi^i\cD\psi^i}}~, \;
B=\frac{A^2}{2(A-2)},\,C=-\frac{A^4}{2(A-2)},\, \nn
&&E=\frac{A^3}{A-2},\,
F=\frac{A^3(A-4)}{6(A-2)^2},\,
G=-\frac{A^5(A-4)}{6(A-2)^2}\,.
\eea
The integral \p{n4action1} with $\Upsilon$ defined by
\p{n1action2}, \p{coeff} provides us with the action for the system
under consideration. 

We can greatly simplify this action. First, the $B$ and 
$C$ terms in \p{n1action2} can be absorbed into the $F$ term. 
All  the remaining $E$, $F$, $G$ terms can be
reduced to the single $A$ term, redefining the superfields
$v^i$ as follows
\be
v^i\rightarrow \phi^i = v^i+ \psi^3 \varepsilon^{ijk}\psi^{j}_t \cD\psi^k
H_1+\psi^3 \psi^i_t H_2+     \varepsilon^{ijk}\psi^{2j}\cD \psi^k H_3 \;,
\ee
where $H_1,\; H_2, \; H_3$ are some functions of $X$. These functions
can be given explicitly, but to know their precise structure 
is of no need for our purposes. The action in terms
of the redefined bosonic superfield $\phi^i$ takes the very simple form
\be\label{action1f}
S_\phi=\int dt d\theta 
\frac{2\xi^i\cD\xi^i}{1+\sqrt{1-4\cD\xi^j\cD\xi^j}}\;,
\quad
\xi^i \equiv \frac{1}{2}\cD\phi^i~.
\ee
By construction, it is guaranteed to be invariant.

Thus we have found the correct Goldstone superfields action
describing the PBGS pattern $N=4\rightarrow N=1$.

Let us end this section by noting that the bosonic core of the action
\p{action1f}
\be
S_\phi^{bos}=\frac{1}{2}\int dt  \left( 1- \sqrt{1-\partial_t \phi^i
  \partial_t \phi^i}  \right)
\ee
is the standard massive $D=4$ particle action in the static gauge.
\vspace{0.3cm}

\noindent{\bf 3. Target space action with one $\kappa$-supersymmetry.}
\hskip 1em
To clarify the situation with $N=4 \rightarrow N=1$ PBGS, 
we construct the target space action which possesses only one
$\kappa$ supersymmetry and reduces to the action \p{action1f} in a
fixed gauge.

We shall deal with the $N=4$ superalgebra \p{susy4}. In accord
with the standard strategy of constructing GS-type
actions for massive superparticles 
(see \cite{{aluk},{town},{town2},{town3},{gauntY}}) we
introduce bosonic $X^0(t),Y^i(t)$ and fermionic  $\Theta (t), \Psi^i (t)$
$d=1$ fields, the coordinates of a target $N=4$ superspace, with the
standard transformation properties under $N=4$ supersymmetry  \p{susy4}
\be\label{GStr}
\delta X^0=-\epsilon \Theta-\epsilon^i\Psi^i ,\;
\delta Y^i=-\epsilon^i\Theta-\epsilon\Psi^i
,\; \delta\Theta=\epsilon ,\;
\delta\Psi^i=\epsilon^i~,
\ee
and construct the invariants $\Pi^0,\Pi^i$
\be\label{GSPi}
\Pi^0={\partial_t X}{}^0+\Theta\partial_t\Theta+
  \Psi^i\partial_t\Psi{}^i~, \;
\Pi^i={\partial_t Y}{}^i-\partial_t \Theta \Psi^i+
 \Theta {\partial_t \Psi}{}^i~.
\ee

After some guess-work, the target sigma-model action invariant
under the global target space supersymmetry \p{GStr}, local
$t$ reparametrizations and one local fermionic $\kappa$
symmetry was found to have the following form
\be\label{GSaction}
S_{gs}=-\int dt \sqrt{\Pi^0\Pi^0 - \Pi^i\Pi^i} - \int dt
     \left( \Theta\partial_t \Theta - \Psi^i\partial_t \Psi{}^i\right) \;.
\ee
The $\kappa$ symmetry transformations are given by
\bea
&& \delta\Theta = \kappa~, \;
\delta\Psi^i= \kappa\; \frac{\Pi^i}{\Pi^0 - \sqrt{\Pi^0\Pi^0
- \Pi^i\Pi^i} } \nn
&& \delta X^0=-\Theta\delta\Theta-\Psi^i\delta\Psi^i ,\;
\delta Y^i=-\Psi^i\delta\Theta-
 \Theta\delta\Psi^i~,\label{GSkappa}
\eea
where $\kappa(t)$ is an arbitrary fermionic gauge parameter. 

The action \p{GSaction} possesses only one $\kappa$ supersymmetry 
and therefore provides a ``space-time''  realization of the 
$N=4\rightarrow N=1$ PBGS phenomenon. 

To prove that there are no any other local fermionic symmetry 
in \p{GSaction} apart from $\kappa$-symmetry \p{GSkappa}, we need to study 
the algebra of the constraints in the
Hamiltonian formalism. We first introduce the einbein $e(t)$ and
rewrite the action \p{GSaction} as
\begin{eqnarray}
&& S_{gs}=\int dt
L=-\int dt\left[ \frac{1}{2e}\left( \Pi^0\Pi^0-\Pi^i\Pi^i\right)+
 \frac{e}{2}\right] \nn
&&-\int dt\left( \Theta\partial_t \Theta-\Psi^i\partial_t \Psi^i\right)\;.
\end{eqnarray}
Then we compute
canonically conjugated variables
\begin{eqnarray}
&&P^0 =-\frac{\Pi^0}{e}\;,\;
P^i =\frac{\Pi^i}{e}\;, \;
P_e=0\;,\;
\Omega =
  \left( \frac{\Pi^0}{e}+1\right)\Theta-
\frac{\Pi^i}{e}\Psi^i\;,\nn
&&\Omega^i=
 \left( \frac{\Pi^0}{e}-1\right)\Psi^i-
 \frac{\Pi^i}{e}\Theta \;.
\end{eqnarray}
The canonical hamiltonian reads
\begin{equation}
H=-{e\over 2}(P^0P^0-P^iP^i-1)~.
\end{equation}
There is one primary bosonic constraint, $P_e$, and 
four fermionic constraints
\begin{equation}
\tau^0=\Omega+ (P^0-1)\Theta+P^i\Psi^i~, \quad
\tau^i=\Omega^i+(P^0+1)\Psi^i+P^i\Theta\;.
\end{equation}
When taking the Poisson bracket of the primary bosonic constraint 
with the canonical
hamiltonian, we obtain the secondary bosonic constraint
\begin{equation}  P^0P^0-P^iP^i=1~.  \end{equation}
We now have to determine which of the fermionic
constraints $\tau^\mu = (\tau^0, \tau^i)$ are first class,
and thus generate  gauge symmetries, and which are second class. 
We compute
the matrix of the Poisson brackets of the fermionic constraints
\begin{equation}
\{\tau^\mu,\tau^\nu\}=C^{\mu\nu}~, \quad
C=2\left(\begin{array}{cc}P^0-1 & \vec P^t \\
\vec P & (P^0+1)\mbox{\bf 1}
\end{array}\right)~,
\end{equation}
where {\bf 1} is the $3\times 3$ unit matrix. 
The eigenvalues of $C$ are easily
computed to be $P^0+1$, $P^0+1$, $P^0+\sqrt{\vec P^2+1}$,
$P^0-\sqrt{\vec P^2+1}$. On the constraint surface, the last of 
these eigenvalues vanishes,
and the other three remain non zero. Thus, there is only one first
class constraint which may be chosen to be
\begin{equation}
\kappa=\tau^0-{1\over P^0+1}\vec P\vec\tau.
\end{equation}
Its Poisson brackets with the constraints read
\begin{equation}
\{\kappa,\tau^0\}={2(P^0P^0-P^iP^i-1)\over P^0+1}~,
\qquad \{\kappa,\tau^i\}=0~.
\end{equation}
This constraint generates the unique local fermionic symmetry
\p{GSkappa} through the Poisson bracket.

In the static gauge the action \p{GSaction} reads
\be\label{GSaction3}
S_{gs}=-\int dt \left[ \sqrt{ \left(1+ 
\Psi^i{\partial_t  \Psi}{}^i\right)^2-
   {\partial_t  Y}{}^i{\partial_t  Y}{}^i }- \Psi^i{\partial_t
\Psi}{}^i\right]~.
\ee
It is straightforward to show that it is related to the 
component form of the action \p{action1f} by a field redefinition.  
\vspace{0.3cm}

\noindent{\bf 4. N=8 $\rightarrow$ N=2 PBGS.} \hskip 1em
To construct a
superparticle model which would exhibit $N=8 \rightarrow N=2$ PBGS
we should, before all, examine how 6 broken
supersymmetries could be realized on a set of $N=2, \;d=1$ superfields.
We succeeded in finding two such realizations.
\vspace{0.3cm}

\noindent {\it Case I.} In the first realization the basic 
set of $N=2,d=1$ superfields
consists of seven bosonic superfields: a general real superfield 
$\Phi$ and
two conjugated triplets of chiral-anti-chiral superfields
${\bar v}{}^i, v^i$:
$$D v^i = \Db {\bar v}{}^i =0\;, \quad i=1,2,3 \;,$$
where
\be
D=\frac{\partial}{\partial \theta}+\frac{1}{2}\tb\partial_{t}, \;
\Db=\frac{\partial}{\partial \tb}+\frac{1}{2}\theta\partial_{t}, \;
\ee
The broken supersymmetry transformations read
\be
\delta v^i = -2\left( \tb - D \Phi\right) \epsilon^i
+\varepsilon^{ijk}{\bar\epsilon}{}^j D {\bar v}{}^k~, \;
\delta \Phi = \frac{1}{2}\left( \epsilon^i D {\bar v}{}^i + 
{\bar\epsilon}{}^i\Db v^i \right)~.
\label{n83}
\ee
Together with the manifest
supersymmetry, they form the algebra with six central charges 
$Z^i,{\bar Z}{}^i$
\bea
&& \left\{ Q,{\bar Q} \right\} = P ,\;
\left\{ S^i,{\bar S}{}^j \right\} = \delta^{ij} P~, \nn 
&& \left\{ Q, S^i \right\} =2 Z^i~, \;
\left\{ {\bar Q}, {\bar S}{}^i \right\} = 2{\bar Z}{}^i\;. \label{n84}
\eea
The fermionic chiral superfields defined by
$$\psi^i = -\frac{1}{2} \Db v^i\; , \;  \bar\psi{}^i = 
\frac{1}{2} D {\bar v}{}^i $$
are transformed under \p{n83} as
\be
\delta \psi^i = \left( 1 - \Db D\Phi\right) \epsilon^i
+\varepsilon^{ijk}{\bar\epsilon}{}^j \Db {\bar \psi}{}^k ,\;
\delta \Phi =\epsilon^i{\bar\psi}{}^i- {\bar\epsilon}{}^i\psi^i  .
\label{n81}
\ee
So they are Goldstone superfields corresponding to
the linear realization of six spontaneously broken supersymmetries
with the parameters $\epsilon^i, {\bar\epsilon}{}^i$. The bosonic
superfields ${\bar v}{}^i, v^i$ are the Goldstone ones associated
with the spontaneously broken central charges transformations.

Once again, the superfield $\Phi$, in accord with its transformation
properties, can be chosen as the Lagrangian density for
this  PBGS pattern. 

To express $\Phi$ in terms of the Goldstone superfields
$\psi^i,{\bar\psi}{}^i$, one can apply the method of ref. 
\cite{{ikap1},{ikap2}}. It basically consists in passing 
to another superfield basis by performing a finite spontaneously 
broken supersymmetry transformation with the Goldstone fermionic 
superfields as the parameters. The redefined superfield $\tilde{\Phi} = 
\Phi + O(\psi, \bar\psi)$ transforms homogeneously and so it can be 
put equal to zero with preserving covariance under all supersymmetries. 
This produces equations allowing one to express $\Phi$ in terms  of 
$\psi^i, \bar\psi^i$. 

The straightforward application of this method to the present case   
yields a rather complicated system of equations.
It can be easily solved in the limit of vanishing fermions,
yielding the static gauge action for a massive particle in a
7-dimensional space-time as the bosonic part of the full
superfield action
\be\label{n85}
S_v^{bos}=\frac{1}{2}\int dt 
\left( 1- \sqrt{1+ \partial_t v^i \partial_t{\bar v}{}^i}
 \right)~.
\ee

The GS formulation for this case
is very similar to the case of $N=4\rightarrow N=1$ PBGS.
We define the standard realization of $N=8$
superalgebra \p{n84} in the superspace with seven bosonic
$X^0,Y^i,{\bar
Y}{}^i$ and eight fermionic $\Theta,{\bar\Theta},\Psi^i,{\bar\Psi}{}^i$
coordinates:
\bea\label{n86}
 \delta \Theta &= &\epsilon~, \; \delta \Psi^i = \epsilon^i~, \;
       \delta Y^i=-2\epsilon^i \Theta~, \nn
\delta X^0&= &-\frac{1}{2}\left( \epsilon\bar\Theta+\bar\epsilon\Theta+
    \epsilon^i{\bar\Psi}{}^i+{\bar\epsilon}{}^i \Psi^i \right)~.
\eea
Defining the invariants
\bea\label{n87}
&&\Pi^0={\partial_t  X}{}^0+\frac{1}{2}\left( 
\Theta {\partial_t {\bar\Theta}}
+{\bar\Theta}{\partial_t \Theta} +\Psi^i{\partial_t {\bar \Psi}}{}^i+
{\bar\Psi}{}^i{\partial_t \Psi}{}^i\right)\;,\nn
&&\Pi^i= {\partial_t  Y}{}^i+2{\Psi^i}{\partial_t \Theta}~, \quad
{\bar\Pi}{}^i \equiv \bar{\left( \Pi^i\right)}~.
\eea
we can construct the unique action
\be\label{n88}
S_{gs}=-\int dt \sqrt{ \Pi^0\Pi^0 - \Pi^i{\bar{\Pi}}{}^i } +
   \int dt \left( \Theta{\bar{\partial_t {\Theta}}} -
\Psi^i {\partial_t {\bar\Psi}}{}^i
   \right)~,
\ee
with two $\kappa$ supersymmetries:
\bea\label{n89}
\delta X^0 &=& -\frac{1}{2}\left( \bar\Theta\delta\Theta+
  \Theta\delta{\bar\Theta} +
    {\bar\Psi}{}^i\delta\Psi^i+\Psi^i\delta{\bar\Psi}{}^i \right)~,\nn
 \delta Y^i &=& -2\Psi^i\delta\Theta~, \nn
 \delta \Psi^i &=& \frac{ \Pi^i\delta{\bar\Theta} }
      {  \Pi^0+\sqrt{  \Pi^0\Pi^0 - \Pi^i{\bar{\Pi}}{}^i } }~.
\eea
The Hamiltonian analysis, which repeats the basic steps of the
analysis in the $N=4\rightarrow N=1$ case, shows that there are
no further gauge fermionic symmetries in the action \p{n88}.

In the static gauge, $ X^0=t, \Theta=0$, the action \p{n88} takes the very
simple form
\bea 
S_{gs} &=& -\int dt \left[ \sqrt{ \left(1+\frac{1}{2}\Psi^i{\partial_t
{\bar\Psi}}{}^i+
\frac{1}{2} {\partial_t \Psi}{}^i{\bar\Psi}{}^i \right)^2 +
  {\partial_t  Y}{}^i{\partial_t {\bar{Y}}}{}^i } \right. \nn
&& \left. +\; \Psi^i{\partial_t {\bar\Psi}}{}^i \right]
\;. \label{n810}
\eea
\vspace{0.3cm}

\noindent{\it Case II.} The second realization of 
$N=8, d=1$ supersymmetry with
six spontaneously broken supersymmetries can be constructed in terms
of general bosonic $N=2$ superfield $\Phi$ and six chiral and anti-chiral
Goldstone fermions  $\left\{ \psi_{\alpha},{\bar \psi}_\alpha, \xi,
\bar\xi\right\}, \alpha=1,2$ \be\label{n811}
\Db \psi_\alpha=\Db \xi=0~, \qquad D {\bar\psi}_\alpha= D {\bar\xi}=0\;,\;
\ee
which form two doublets and two singlets with respect to
$SO(2)$ automorphism group. The appropriate closed set of
the broken supersymmetry transformations reads
$$\delta\xi =\left(1+\Db D\Phi\right) \nu +
 \varepsilon_{\alpha\beta}{\bar\mu}_\alpha \Db{\bar\psi}_\beta~, 
$$
$$
\delta\Phi= {\bar\nu} \xi-\nu{\bar\xi}-
   {\bar\mu}_\alpha\psi_\alpha+\mu_\alpha{\bar\psi}_\alpha\,
$$
\be\label{n812}
\delta\psi_\alpha =\varepsilon_{\alpha\beta}\left( \bar\nu
   \Db{\bar\psi}_\beta +{\bar\mu}_\beta \Db{\bar\xi}\right)+
 \left(1-\Db D\Phi\right)\mu_\alpha \; .
\ee

To reveal the underlying central-charges extended supersymmetry
algebra and to gain physical bosonic fields, we need to pass as before
to bosonic superfields. The minimal realization amounts
to introducing two real scalar superfields $u_\alpha$:
\be\label{n814}
\psi_\alpha=-\frac{1}{2}\Db u_\alpha~, 
\quad {\bar\psi}_\alpha=\frac{1}{2} D u_\alpha\;.
\ee
To learn what kind of ``prepotential'' one should introduce
for the remaining Goldstone superfield $\xi$, let us examine
the relation between $U(1)$ charges of spinor superfields which
follows from \p{n812}
\be\label{n813}
q_{\xi}=-2q_{\psi}-q_D\;.
\ee
Here $q_D$ is the $U(1)$ charge of the covariant derivative $D$
($q_D = -1$ if one ascribes the charge $+1$ to $\theta$).
{}From this relation and eq. \p{n814} it follows that
the only  way to introduce the bosonic
superfield $v$ for $\xi$ is to choose it complex and having
the $U(1)$ charge $-2q_D$
\be\label{n815}
\xi=-\frac{1}{2}\Db v~, \quad {\bar\xi} = \frac{1}{2}D {\bar v}~,
\quad Dv = \Db{\bar v} =0 \;.
\ee
In terms of $v, {\bar v} $ the
supersymmetry transformations become:
\bea\label{n816}
\delta v &=&-2\left(\tb+ D\Phi\right) \nu +
\varepsilon_{\alpha\beta}{\bar\mu}_\alpha D u_\beta~, \nn
\delta u_\alpha&=& \varepsilon_{\alpha\beta}\left(
 {\bar\nu} D u_\beta+\nu \Db u_\beta +
 {\bar\mu}_\beta D {\bar v} +\mu_\beta \Db v\right)
    \nn
&&   
+ 2\left(\theta+ \Db\Phi\right){\bar\mu}_\alpha -
2\left({\bar\theta}- D\Phi\right)\mu_\alpha\; , \nn
\delta\Phi &=& -\frac{1}{2}\left( \nu D{\bar v}+{\bar\nu}\Db v-
   {\bar\mu}_\alpha \Db u_\alpha- \mu_\alpha D u_\alpha\right)\; .
\eea
Denoting the generators of the broken supersymmetry by
$S_\alpha, {\bar S}_\alpha$ and $S, {\bar S}$, and the generators of
the manifest $N=2$ supersymmetry by $Q,{\bar Q}$, one can write the
full supersymmetry algebra pertinent to this case as
\bea\label{n817}
&& \left\{ Q,{\bar Q} \right\}= \left\{ S,{\bar S} \right\}=P,\;
\left\{ S_\alpha,{\bar S}_\beta \right\}=\delta_{\alpha,\beta}P, \;
 \nn
&& 
\left\{ Q,{\bar S} \right\}=2{\bar Z}, \;
\left\{ {\bar Q}, S \right\}=2Z,\;
 \left\{ Q, {\bar S}_\alpha \right\}=2Z_\alpha, \;
  \nn
&&
\left\{ {\bar Q}, S_\alpha \right\}=2Z_\alpha,\,
\left\{ S, {\bar S}_\alpha \right\}=
 2\varepsilon_{\alpha\beta}Z_\beta,\,
 \left\{ {\bar S},S_\alpha \right\}=
 2\varepsilon_{\alpha\beta}Z_\beta\,.
\eea

Once again, we can take the superfield $\Phi$ as the Lagrangian
density. To covariantly express $\Phi$ in terms of
the Goldstone fermions or Goldstone bosons, we may again stick to 
the general method of \cite{ikap1,ikap2}. However it gives
rather complicated equations which we for the time being were unable to
solve. We could try to find at least the bosonic part of the
action. We get the following 
quartic equation for the bosonic part of the action which we denote by $X$:
\be\label{n820}
(X^2-X+a)(X^2+a-1)+2D\xi\Db{\bar\xi} = 0 \; ,
\ee
where $$a= D\xi\Db{\bar\xi}+  \Db \psi_\alpha D{\bar\psi}_\alpha  \;.$$
The general solution of this equation exists (we require it to 
vanish in the limit when all fields are put equal to zero),
but it looks not too illuminating to present it here.
In the two limits,  $\psi_\alpha=0$ or $\xi=0$, it takes the familiar
form of the static gauge actions of massive particles moving on
some 3-dimensional target manifolds
\bea\label{n822}
S_v^{bos} &= &\frac{1}{2}\int dt 
\left( 1-\sqrt{1+ {\partial_t  v}{\partial_t {\bar v} }}
        \right)~, \nn
S_u^{bos} &= &\frac{1}{2}\int dt \left( 1-\sqrt{ 1+
       {\partial_t  u}_\alpha {\partial_t  u}_\alpha}
       \right)\;.
\eea
In the generic case there is a non-trivial cross-interaction between
the bosonic fields appearing in \p{n822}. It can hopefully be interpreted
in terms of intersection of the trajectories of two
different superparticles, with the physical worldline
scalar multiplets represented by the superfields $u_\alpha$ and
$v, \bar v $, respectively.

The fact that the bosonic part of the action cannot be written in
the standard static gauge Nambu-Goto form seriously obscures 
the construction of the GS
formulation for this case. 
We believe that the better understanding of this case
would be  helpful for studying  the $1/4$ PBGS systems with
higher-dimensional worldvolumes. 
\vspace{0.3cm}

\noindent{\bf 5. Conclusions.}\hskip 1em
In this paper we presented, for the first time, the manifestly worldline
supersymmetric superparticle actions exhibiting hidden spontaneously
broken supersymmetries the number of which is four times the number of
the linearly realized manifest ones. We treated in detail the case
of $N=4 \rightarrow N=1$ partial breaking and discussed some basic features
of the more complicated $N=8 \rightarrow N=2$ case.  
The common unusual feature of the
superparticle systems considered is that their space-time
interpretation is possible only within the superspaces corresponding
to higher-dimensional supersymmetries with tensorial central charge
generators. 
It would be of interest to understand
whether this is the general property of systems with fractional
PBGS. 
\vspace{0.3cm}

\noindent{\bf Acknowledgements.}\hskip 1em
We thank A.~Kapustnikov, J.~Lukierski, D.~L\"{u}st, A.~Pashnev,
P.~Pasti,
C.~Preitschopf, D.~Sorokin,  M.~Tonin and B.~Zupnik for many useful
discussions. E.I. and S.K. are grateful to Organizers of XIV-th
Max Born Symposium for inviting them to present this talk.
This work was supported in part
by the PICS
Project No. 593, RFBR-CNRS Grant No. 98-02-22034, 
RFBR Grant No. 99-02-18417,
Nato Grant No. PST.CLG 974874 and INTAS Grants INTAS-96-0538, INTAS-96-0308.

\end{document}